\def\year{2021}\relax
\documentclass[letterpaper]{article} 
\usepackage{aaai21}  
\usepackage{times}  
\usepackage{helvet} 
\usepackage{courier}  
\usepackage[hyphens]{url}  
\usepackage{graphicx} 
\usepackage{natbib}  
\usepackage{caption} 
\nocopyright
\title{The \emph{Fair}Ceptron:\\ A Framework for Measuring Human Perceptions of Algorithmic Fairness}
\author{
    Georg Ahnert,\textsuperscript{\rm 1} Ivan Smirnov,\textsuperscript{\rm 1} Florian Lemmerich,\textsuperscript{\rm 1} Claudia Wagner,\textsuperscript{\rm 2} Markus Strohmaier\textsuperscript{\rm 1,\rm 2}
}
\affiliations{
    
    \textsuperscript{\rm 1}RWTH Aachen University
    \textsuperscript{\rm 2}GESIS - Leibniz Institute for the Social Sciences
    
    georg.ahnert@rwth-aachen.de, \{ivan.smirnov, florian.lemmerich, markus.strohmaier\}@cssh.rwth-aachen.de, claudia.wagner@gesis.org
}

\begin{document}

\maketitle

\begin{abstract}
Measures of algorithmic fairness often do not account for human perceptions of fairness that can substantially vary between different sociodemographics and stakeholders. The \textit{Fair}Ceptron framework is an approach for studying perceptions of fairness in algorithmic decision making such as in ranking or classification. It supports (i) studying human perceptions of fairness and (ii) comparing these human perceptions with measures of algorithmic fairness. The framework includes fairness scenario generation, fairness perception elicitation and fairness perception analysis. We demonstrate the \textit{Fair}Ceptron framework by applying it to a hypothetical university admission context where we collect human perceptions of fairness in the presence of minorities. An implementation of the \textit{Fair}Ceptron framework is openly available\footnote{https://github.com/cssh-rwth/fairceptron}, and it can easily be adapted to study perceptions of algorithmic fairness in other application contexts. We hope our work paves the way towards elevating the role of studies of human fairness perceptions in the process of designing algorithmic decision making systems.
\end{abstract}

\section{Motivation}
Considering fairness in algorithmic decision-making poses an important challenge~\cite{chouldechova2020snapshot}. Different definitions of algorithmic fairness have been proposed, including individual measures~\cite{dwork2012fairness}, as well as group based measures for both classification~\cite{friedler2019comparative} and ranking decisions~\cite{yang2017measuring}. In general, algorithms trade accuracy and fairness~\cite{kearns2019ethical}, and group-based fairness measures cannot be simultaneously equalized over all groups~\cite{chouldechova2017prediction}. Thus, normative decisions must be made.

One way of approaching these decisions is through an analysis of what is perceived as fair, involving the target population of a deciding algorithm in its creation. This could increase the acceptance of algorithmic decision making~\cite{awad2018moral}. Involvement also benefits procedural fairness, often the most important contributor to overall fairness perception~\cite{ambrose2015overall}. Previous research investigated perceptions of algorithmic fairness~\cite{saxena2019fairness, srivastava2019mathematical, harrison2020empirical}, but focused on classification and predominantly optimal decisions.

Psychological research suggests that fairness perceptions are influenced by social context~\cite{engstrom2020justification}. It was found that fairness perception differs between genders~\cite{dulebohn2016gender}, cultures~\cite{blake2015ontogeny}, and people of different personality traits~\cite{truxillo2006field, wiesenfeld2007more}. These differences are currently not accounted for in fairness measures commonly used in computer science.

In this paper we present the \textit{Fair}Ceptron framework for studying fairness perceptions. It allows to study classification and ranking decisions that do not necessarily optimize for a single fairness measure. With the \textit{Fair}Ceptron, obligatory trade-offs between accuracy and multiple fairness measures can be investigated, and the nature of the relationships between fairness perceptions and fairness measures can be determined. An implementation is available as open source\footnotemark[1] and built for easy deployment and adaptation to different study contexts.

\begin{figure*}[t!]
\centering
\includegraphics[width=0.97\textwidth]{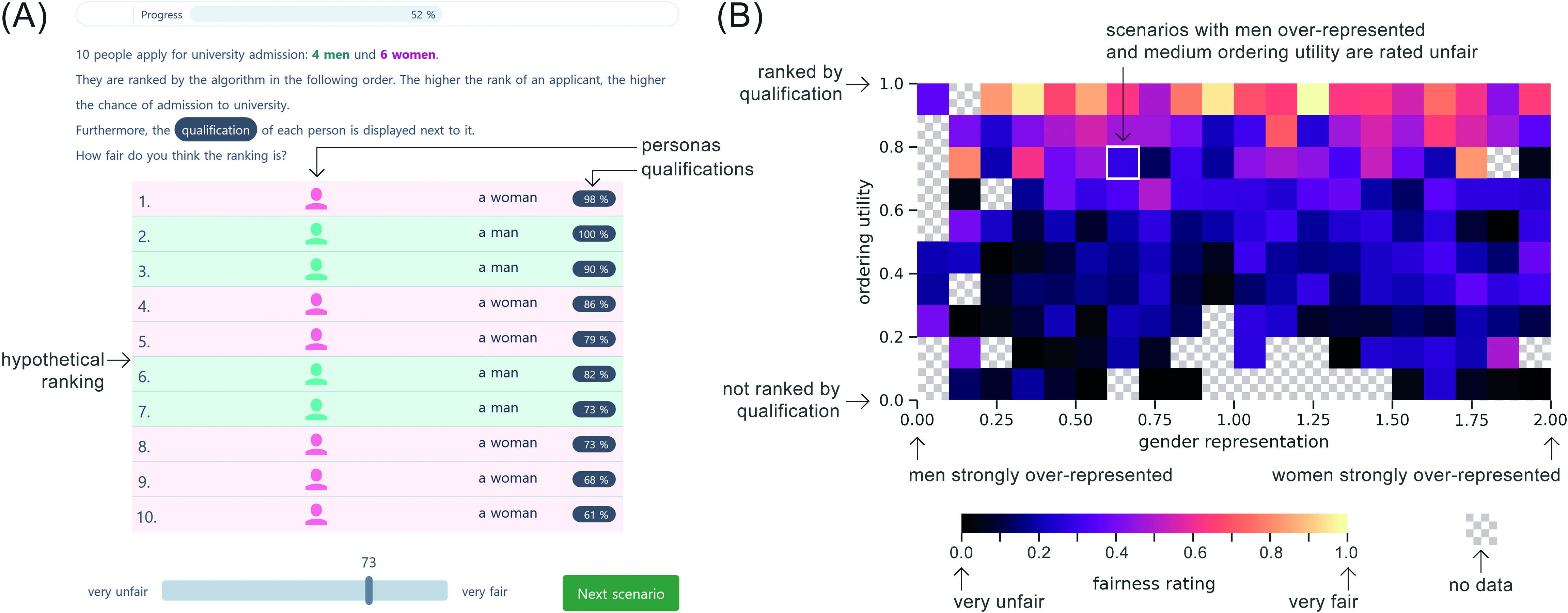} 
\caption{\textbf{(A) A \textit{Fair}Ceptron ranking scenario.} Participants are shown an algorithmic ranking scenario. They rate perceived fairness of the scenario on a visual analogue scale. In addition to ranking, classification scenarios are also supported.\\
\textbf{(B) Perceptions of fairness across different ranking scenarios.} All scenarios are binned by ordering utility~\cite{zehlike2017fa} and gender representation (adapted from \citeauthor{yang2017measuring}~\citeyear{yang2017measuring}). Participants were mainly influenced by ordering utility. Higher ratings for over-representation of women vs. men can be seen in scenarios with ordering utility $< 0.7$.}
\label{main figure}
\end{figure*}

\section{The \textit{Fair}Ceptron Framework}
The \textit{Fair}Ceptron framework consist of three components: (i) the generation of fairness scenarios according to a prespecified algorithm, (ii) presentation of scenarios to survey participants and collecting their subjective fairness rating, and (iii) analysis of responses that takes into account characteristics of scenarios, e.g. group sizes, and characteristics of users, e.g. sociodemographics or attitudes. The \textit{Fair}Ceptron framework can be implemented in various ways, in this paper we present one particular implementation. 

\subsubsection{Fairness scenario generation} Algorithmic ranking and classification scenarios are generated that consist of personas of two or more groups that can optionally have a second, numeric attribute associated to them. We provide simple code examples for scenario generation in Python. The scenarios are generated as all possible selections from / permutations of $n$ personas, in which personas within a group are selected / ranked by qualification. The scenarios are clustered along multiple measures of algorithmic fairness, ensuring that each participant later receives a variety of scenarios, while maximizing the total number of scenarios that are tested.

\subsubsection{Fairness perception elicitation} Participants take part through a responsive, universal web application as shown in Fig.~\ref{main figure} (A). For each new participant, the application selects one random fairness scenario from each pre-defined cluster of scenarios, and then shuffles the selected scenarios. For every scenario, a description and an illustration is shown. The participants rate each scenario on an initially blank visual analogue scale (VAS) from \textit{very unfair} to \textit{very fair}. A dynamic indicator is added to the VAS to improve accuracy with minimal additional bias~\cite{matejka2016effect}. The time to answer, and the uncertainty in answering, measured as the sum of differences of non-final ratings, are stored alongside the final answer. Sociodemographics and attitudes can also be elicited.

\subsubsection{Fairness perception analysis} The obtained data can be exported from MongoDB in CSV or JSON format. We provide evaluation examples written with common Python frameworks for the above listed analyses. Heatmaps that compare fairness ratings on scenarios group by two distinct measures can easily be generated, as shown in Fig.~\ref{main figure} (B).

\section{Demonstration}
For demonstration purposes, we applied the \textit{Fair}Ceptron framework using a voluntary response sample of 136 people. The hypothetical scenarios concern a university admission process. All scenarios displayed 10 female / male student applicants with associated \textit{qualification scores}. Each participant was asked to rate 10 classification and 10 ranking scenarios. Additionally, participants filled in additional questions about their demographics, their attitudes towards deciding machines (adapted from \citeauthor{awad2018moral} \citeyear{awad2018moral}), and took a big-five personality short test~\cite{rammstedt2007measuring}.

Fig.~\ref{main figure} (B) illustrates the fairness perceptions aggregated from the ranking scenarios of the \textit{Fair}Ceptron study. In general, participants rated scenarios according to their ordering utility. The highlighted exemplary bin is rated unfair on average, with scenarios that partially violate qualification order and in which men are over-represented. Ratings differ by participant gender and political orientation, in particular the acceptance of over-representing female personas. These findings only serve for illustration and are obtained from a non-representative population. The demo at ICWSM will include a walk-through over scenario generation, perception elicitation, and analysis.

\textit{Fair}Ceptron studies can easily be deployed with little efforts building upon the existing implementation. The framework allows to investigate whether fairness perceptions depend on domains (e.g. education, medicine, finance), sociodemographics (e.g. gender, occupation) or the stakes involved (high- vs low-stakes decisions). The results obtained from \textit{Fair}Ceptron studies could empirically inform the selection and evaluation of fairness measures in real world settings. We hope our framework represents a stepping stone towards a future, in which the people subjected to algorithmic decision making are contributing in its design process, and in which algorithmic notions of fairness are subjected to empirical studies of human perceptions of fairness before implementation and roll-out.

In summary, we present a framework for studying perceptions of fairness in algorithmic decision making such as in ranking or classification that includes fairness scenario generation, fairness elicitation and fairness perception analysis steps. Our implementation of the framework is available on GitHub as open source.

\bibliography{main.bib}

\end{document}